\documentclass[aps,preprint,showpacs,preprintnumbers,eqsecnum,amsmath,amssymb]{revtex4}

\usepackage{graphics,epsfig}
\usepackage{graphicx}
\usepackage{dcolumn}
\usepackage{bm}

\begin{document}

\title{Dirac Quasinormal modes of Schwarzschild black hole}
\author{Jiliang Jing} \email{jljing@hunnu.edu.cn}
\affiliation{ Institute of Physics and  Department of Physics, \\
Hunan Normal University,\\ Changsha, Hunan 410081, P. R. China }

 \baselineskip=0.65 cm

\vspace*{0.2cm}
\begin{abstract}
\vspace*{0.2cm}

The quasinormal modes (QNMs) associated with the decay of Dirac
field perturbation around a Schwarzschild black hole is
investigated by using continued fraction and Hill-determinant
approaches. It is shown that the fundamental quasinormal
frequencies become evenly spaced for large angular quantum number
and the spacing is given by $\omega_{\lambda+1}-
\omega_{\lambda}=0.38490-0.00000i$. The angular quantum number has
the surprising effect of increasing real part of the quasinormal
frequencies, but it almost does not affect imaginary part,
especially for low overtones. In addition, the quasinormal
frequencies also become evenly spaced for large overtone number
and the spacing for imaginary part is
$Im(\omega_{n+1})-Im(\omega_n)\approx -i/4M$ which is same as that
of the scalar, electromagnetic, and gravitational perturbations.

\end{abstract}

 \vspace*{1.5cm}

\maketitle

\section{introduction}

Motivated by the beliefs that one can directly identify a black
hole existence by comparing QNMs with the gravitational waves
observed in the universe, and the study of QNMs may lead to a
deeper understanding of the thermodynamic properties of black
holes in loop quantum gravity \cite{Hod} \cite{Dreyer}, as well as
the QNMs of anti-de Sitter black holes have a direction
interpretation in terms of the dual conformal field theory
\cite{Maldacena} \cite{Witten} \cite{Kalyana}, the investigation
of QNMs in black hole spacetimes becomes appealing recent years.
Since Regge and Wheeler \cite{Regge} first presented the idea of
the QNMs and Chandrasekhar and Detweller \cite{Chand75} worked out
the numerically calculation of the QNMs for the Schwarzschild
black hole, a great deal of effort has been contributed to compute
the QNMs of the black holes \cite{Chan}-\cite{Konoplya1} for
fields of integer spin, such as the scalar, electromagnetic and
gravitational perturbations.

On the contrary, the study of the QNMs of the Dirac field is very
limited \cite{Cho}\cite{Zhidenko}\cite{Jing1}. Cho \cite{Cho}
studied the QNMs of the Dirac field of the Schwarzschild black
hole with the help of the third-order WKB method, and  Zhidenko
\cite{Zhidenko} extended the study to the case of the
Schwarzschild-de Sitter black hole by using the sixth-order WKB
method. We \cite{Jing1} investigated the Dirac QNMs of the
Reissner-Nordstr\"om de Sitter black hole using the P\"oshl-Teller
potential approximation. Nevertheless, in these papers the study
of the Dirac QNMs were limited by using WKB or P\"oshl-Teller
potential approximative methods. The reason to obstruct study of
the Dirac QNMs is that for static black holes the standard
wave-equations
\begin{eqnarray}\label{sta}
\left(\frac{d^2}{d r_*^2}+\omega^2\right)Z_{\pm}=V_{\pm}Z_{\pm}
\end{eqnarray}
possesses the special potentials
 \begin{eqnarray}
V_{\pm}=\lambda^2\frac{ \Delta }{r^4}\pm \lambda \frac{d}{d
r_*}\frac{\sqrt{\Delta}}{r^2}
 \end{eqnarray}
which are the functions of $\sqrt{\Delta}$
\cite{Cho}\cite{Jing1}\cite{Jing2}, where $\Delta$ is a function
related to metric, say $\Delta=r^2-2Mr$ for the Schwarzschild
black hole. So, it is very hard to calculate Dirac QNMs by using
the known numerical methods (except WKB or P\"oshl-Teller
approximative methods) since we have to expand the potentials as a
series at boundary of the field, such as at the event horizon.
However, Castello-Branco {\textit {et al}} \cite{Branco} evaluated
the Dirac QNMs of the Schwarzschild black hole by using the
continued fraction approach in which they expanded the wave
function as a series of $\sqrt{\Delta}/r$, and they found that the
spacing for imaginary part is not $-i/4M$, as it takes place for
scalar, electromagnetic, and gravitational perturbations, but
$-i/8M$.

Recently, we \cite{Jing3}\cite{Jing4} found that the wave
functions and potentials of the Dirac field in the static
spacetimes can be expressed as new forms, and the new wave
functions are related to the standard wave functions $Z_{\pm}$ in
Eq. (\ref{sta}) in a simple way \cite{Jing3}. Starting from the
new wave functions and potentials, we \cite{Jing3}\cite{Jing4}
showed that the Dirac QNMs of the Schwarzschild anti-de Sitter and
Reissner-Norstr\"om anti-de Sitter black holes can easily be
worked out by using Horowitz-Hubeny approach \cite{Horowitz}. We
have checked that the new potentials also present correct
quasinormal frequencies for the scalar and gravitational
perturbations in the Schwarzschild black hole.

The main purpose of this paper is to study Dirac QNMs of the
Schwarzschild black hole by using the new kinds of wave functions
and potentials with the help of the continued fraction and
Hill-determinant approaches. The results obtained here are
different from Castello-Branco's one \cite{Branco} (If taking
$M=1$, we can only obtain results $n=0, 2,...$ in Ref.
\cite{Branco}. That is to say we do not find ``specific modes".).
We are sure that we do not miss any quasinormal frequency because
the Hill-determinant approach gives all roots for $n\leq 15$.

The organization of this paper is as follows. In Sec. 2 the
decoupled Dirac equations and corresponding wave equations in the
Schwarzschild spacetime are obtained by using Newman-Penrose
formalism. In Sec. 3 the numerical approach to computing the Dirac
QNMs is introduced. In Sec. 4 the numerical results for the Dirac
QNMs in the Schwarzschild black hole are presented. The last
section is devoted to a summary.

\section{Dirac equations in the Schwarzschild spacetime}

The Dirac equations \cite{Page} are
\begin{eqnarray}
   &&\sqrt{2}\nabla_{BB'}P^B+i\mu \bar{Q}_{B'}=0, \nonumber \\
   &&\sqrt{2}\nabla_{BB'}Q^B+i\mu \bar{P}_{B'}=0,
\end{eqnarray}
where $\nabla_{BB'}$ is covariant differentiation, $P^B$ and $Q^B$
are the two-component spinors representing the wave function, $
\bar{P}_{B'}$ is the complex conjugate of $P_{B}$, and $\mu $ is
the particle mass. In the Newman-Penrose formalism \cite{Newman}
the equations become
\begin{eqnarray}\label{np}
   &&(D+\epsilon-\rho )P^0+
   (\bar{\delta}+\pi-\alpha )P^1=2^{-1/2}i\mu  \bar{Q}^{1'},\nonumber \\
   &&(\triangle+\mu -\gamma )P^1+
   (\delta+\beta -\tau )P^0=-2^{-1/2}i\mu
    \bar{Q}^{0'}, \nonumber\\
   &&(D+\bar{\epsilon}-\bar{\rho} )\bar{Q}^{0'}+
   (\delta+\bar{\pi}-\bar{\alpha} )\bar{Q}^{1'}=-2^{-1/2}i\mu  P^{1},\nonumber \\
  &&(\triangle+\bar{\mu} -\bar{\gamma} )\bar{Q}^{1'}+
   (\bar{\delta}+\bar{\beta} -\bar{\tau} )\bar{Q}^{0'}=2^{-1/2}i\mu
   P^{0}.
\end{eqnarray}
For the Schwarzschild spacetime
\begin{eqnarray} \label{metric}
ds^2=\left(1-\frac{2 M}{r}\right) dt^2-\frac{1}{1-\frac{2
M}{r}}dr^2-r^2(d\theta^2+sin^2\theta d\varphi^2),
\end{eqnarray}
where M represents the black hole mass,  the null tetrad can be
taken as
\begin{eqnarray}
  &&l^\mu=(\frac{r^2}{\Delta}, ~1, ~0, ~0 ), \nonumber \\
  &&n^\mu=\frac{1}{2}(1, ~-\frac{\Delta}{r^2}, ~0, ~0)\nonumber \\
  &&m^\mu=\frac{1}{\sqrt{2} r}\left(0, ~0, ~1, \frac{i}{sin\theta}\right),
\end{eqnarray}
with
\begin{eqnarray}
\Delta=r^2-2M r.
\end{eqnarray}
Then, if we set
\begin{eqnarray}
&&P^0=\frac{1}{r}{\mathbb{R}}_{-1/2}(r)S_{-1/2}(\theta)
e^{-i(\omega t-\bar{m}\varphi)}, \nonumber \\
&&P^1={\mathbb{R}}_{+1/2}(r)S_{+1/2}(\theta)
e^{-i(\omega t-\bar{m}\varphi)}, \nonumber \\
&&\bar{Q}^{1'}={\mathbb{R}}_{+1/2}(r)S_{-1/2}(\theta)
e^{-i(\omega t-\bar{m}\varphi)}, \nonumber \\
&&\bar{Q}^{0'}=-\frac{1}{r}{\mathbb{R}}_{-1/2}(r)S_{+1/2}(\theta)e^{-i(\omega
t-\bar{m}\varphi)},
\end{eqnarray}
where $\omega$ and $\bar{m}$ are the energy and angular momentum
of the Dirac particles, after tedious calculation Eq. (\ref{np})
can be simplified as
\begin{eqnarray}\label{dd2}
&&\sqrt{\Delta}{\mathcal{D}}_0 {\mathbb{R}}_{-1/2}=(\lambda+i\mu
r)\sqrt{\Delta} {\mathbb{R}}_{+1/2}, \\
\label{dd3}&&\sqrt{\Delta}{\mathcal{D}}_0^{\dag}
(\sqrt{\Delta}{\mathbb{R}}_{+1/2})=(\lambda-i\mu r) {\mathbb{R}}_{-1/2},\\
&&{\mathcal{L}}_{1/2} S_{+1/2}=-\lambda S_{-1/2}, \label{aa1}\\
&&{\mathcal{L}}_{1/2}^{\dag} S_{-1/2}=\lambda S_{+1/2}.\label{aa2}
\end{eqnarray}
with
 \begin{eqnarray}
 &&{\mathcal{D}}_n=\frac{\partial}{\partial r}-\frac{i K}
 {\Delta}+\frac{n}{\Delta}\frac{d \Delta}{d r},\nonumber \\
 &&{\mathcal{D}}^{\dag}_n=\frac{\partial}{\partial r}+\frac{i K}
 {\Delta}+\frac{n}{\Delta}\frac{d \Delta}{d r},\nonumber \\
 &&{\mathcal{L}}_n=\frac{\partial}{\partial \theta}+\frac{\bar{m}}{\sin \theta }
 +n\cot \theta,\nonumber \\
 &&{\mathcal{L}}^{\dag}_n=\frac{\partial}{\partial \theta}-\frac{\bar{m}}{\sin \theta }
 +n\cot \theta, \nonumber \\
 &&K=r^2\omega.\label{ld}
 \end{eqnarray}

We can eliminate $S_{+1/2}$ (or $S_{-1/2}$) from Eqs. (\ref{aa1})
and (\ref{aa2}) and obtain
\begin{eqnarray}\label{ang}
\left[\frac{1}{sin\theta}\frac{d}{d
\theta}\left(sin\theta\frac{d}{d\theta}\right)-\frac{\bar{m}^2+2\bar{m}scos\theta+s^2cos^2\theta}
{sin^2\theta}+s+A_s\right]S_s=0,
\end{eqnarray}
here and hereafter we take $s=+1/2$ for the case $S_{+1/2}$
(${\mathbb{R}}_{+1/2}$) and $s=-1/2$ for $S_{-1/2}$ (
${\mathbb{R}}_{- 1/2}$), and $A_{+1/2}=\lambda^2-2s$ and
$A_{-1/2}=\lambda^2$.  The angular equation (\ref{ang}) can be
solved exactly and $A_s=(l-s)(l+s+1)$, where $l$ is the quantum
number characterizing the angular distribution. So, for both cases
$s=+1/2$ and $s=-1/2$ we have
 \begin{eqnarray}
\lambda^2=\left(l+\frac{1}{2}\right)^2.
 \end{eqnarray}

We will focus our attention on the massless case in this paper.
Therefore, we can eliminate ${\mathbb{R}}_{-1/2}$ (or
$\sqrt{\Delta}{\mathbb{R}}_{+1/2}$) from Eqs. (\ref{dd2}) and
(\ref{dd3}) to obtain a radial decoupled Dirac equation for
$\sqrt{\Delta} {\mathbb{R}}_{+1/2}$ (or ${\mathbb{R}}_{-1/2}$).
Then, Introducing an usual tortoise coordinate
 \begin{eqnarray}
dr_*=\frac{r^2}{\Delta} dr
  \end{eqnarray}
 and resolving the equation in the form
 \begin{eqnarray} \label{Rphi}
&&{\mathbb{R}}_{s}=\frac{\Delta^{-s/2}}{r} \Psi_s,
 \end{eqnarray}
we obtain the wave equation
\begin{eqnarray}\label{wave}
\frac{d \Psi_s }{d r_*^2}+(\omega ^2-V_s )\Psi_s =0,
\end{eqnarray}
where
\begin{eqnarray}\label{Poten}
V_s=-\frac{\Delta}{4 r^2}\frac{d}{d
r}\left[r^2\frac{d}{dr}\left(\frac{\Delta}{r^4}\right)\right]+\frac{s^2
r^4}{4}\left[\frac{d}{d
r}\left(\frac{\Delta}{r^4}\right)\right]^2+is \omega r^2\frac{d}{d
r}\left(\frac{\Delta}{r^4}\right)+\frac{\lambda^2 \Delta}{r^4}.
\end{eqnarray}
Although the form of the potential is more complicated than
standard wave-equations $\left(\frac{d^2}{d
r_*^2}+\omega^2\right)Z_{\pm}=V_{\pm}Z_{\pm}$ with the potentials
$V_{\pm}=\lambda^2\frac{ \Delta }{r^4}\pm \lambda \frac{d}{d
r_*}\frac{\sqrt{\Delta}}{r^2}$ \cite{Cho}\cite{Jing1}\cite{Jing2},
we will see that we can easily work out the Dirac quasinormal
frequencies of the Schwarzschild black hole from Eqs. (\ref{wave})
and (\ref{Poten}) by using continued fraction and Hill-determinant
approaches.

\section{Numerical Approaches to Computing Dirac Quasinormal Modes}

For the Schwarzschild black hole, the QNMs are defined to be modes
with purely ingoing waves at the event horizon and purely outgoing
waves at infinity \cite{Chand75}. Then, the boundary conditions on
wave function $\Psi_s$ at the horizon $(r=r_+)$ and infinity
$(r\rightarrow +\infty)$ can be expressed as
 \begin{eqnarray}
 \label{Bon}
\Psi_s  \sim \left\{
\begin{array}{ll} (r-r_+)^{-\frac{s}{2}-\frac{i\omega}{2\kappa}} &
~~~~r\rightarrow r_+, \\
     r^{-s+i\omega}e^{i\omega r} & ~~~~     r\rightarrow +\infty,
\end{array} \right.
 \end{eqnarray}
where $\kappa$ is the surface gravity of the black hole.

A solution to Eq. (\ref{wave}) that has the desired behaviour at
the boundary can be written in the form
 \begin{eqnarray}\label{expand}
 \Psi_s=r^{-\frac{s}{2}+2i\omega}(r-r_+)^{-\frac{s}{2}
 -\frac{i\omega}{2\kappa}}e^{i\omega r}\sum_{m=0}^{\infty}
 a_m\left(\frac{r-r_+}{r}\right)^{m}
 \end{eqnarray}
If we take $r_+=1$, the sequence of the expansion coefficient
$\{a_m: m=1,2,....\}$ is determined by a three-term recurrence
relation staring with $a_0=1$:
 \begin{eqnarray} \label{rec}
 &&\alpha_0 a_1+\beta_0 a_0=0, \nonumber \\
 &&\alpha_m a_{m+1}+\beta_m a_m+\gamma_m a_{m-1}=0,~~~m=1,2,....
 \end{eqnarray}
The recurrence coefficient $\alpha_m$, $\beta_m$, and $\gamma_m$
are given in terms of $m$ and the black hole parameters by
\begin{eqnarray}
 &&\alpha_m==m^2+(C_0+1)m+C_0, \nonumber \\
 &&\beta_m=-2m^2+(C_1+2)m+C_3, \nonumber  \\
 &&\gamma_m=m^2+(C_2-3)m+C_4-C_2+2,
 \end{eqnarray}
with
\begin{eqnarray}
 &&C_0=1-s-2i\omega, \nonumber \\
 &&C_1=-4+8i\omega,  \nonumber  \\
 &&C_2=s+3-4i\omega, \nonumber \\
 &&C_3=8\omega^2-s-1+4i\omega-\lambda^2, \nonumber \\
 &&C_4=s+1-4\omega^2-(2s+4)i\omega.
 \end{eqnarray}

The three terms recurrence relation (\ref{rec}) can be used to
find quasinormal frequencies by using both the continued fraction
and Hill-determinant methods which are introduced in the
following.

\vspace*{0.4cm}

 {\bf [The continued fraction approach]:} The ratio
of the series coefficients $\{a_m: m=1, 2, ...\}$ is finite and
can be determined in two ways:
 \begin{eqnarray}\label{an}
 \frac{a_{m+1}}{a_m}&=&\frac{\gamma_m}{\alpha_m}\frac{\alpha_{m-1}}{\beta_{m-1}
 -}\frac{\alpha_{m-2}\gamma_{m-1}}{\beta_{m-2}-}\frac{\alpha_{m-3}
 \gamma_{m-2}}{\beta_{m-3}-}...
 -\frac{\beta_m}{\alpha_m}\nonumber
\\&=&\frac{-\gamma_{m+1}}{\beta_{m+1}
 -}\frac{\alpha_{m+1}\gamma_{m+2}}{\beta_{m+2}-}\frac{\alpha_{m+2}
 \gamma_{m+3}}{\beta_{m+3}-}...
 \end{eqnarray}
Using Eqs. (\ref{rec}) and (\ref{an}) we can obtain an equality
between two continued fractions, one of infinite length, and the
other finite:
\begin{eqnarray}\label{ann}
 \left[\beta_m-\frac{\alpha_{m-1}\gamma_m}{\beta_{m-1}-}
 \frac{\alpha_{m-2}\gamma_{m-1}}{\beta_{m-2}-}...
 \frac{\alpha_0\gamma_1}{\beta_0}\right]=
 \left[\frac{\alpha_m\gamma_{m+1}}{\beta_{m+1}-}
 \frac{\alpha_{m+1}\gamma_{m+2}}{\beta_{m+2}-}
 \frac{\alpha_{m+2}\gamma_{m+3}}{\beta_{m+3}-}...\right],
 ~~(m=1,2...).
 \end{eqnarray}
This leads to a simple method to find quasinormal
frequencies--define a function which returns the value of the
continued fraction for an initial guess at the frequency. Then,
use a root finding routine to find the zeros of this function in
the complex plane. The frequency for which this happens is a
quasinormal frequency. The $n$th quasinormal frequency is usually
found to be numerically the most stable root of the $n$th
inversion \cite{Leaver}.

The success of the root-finding technique depends greatly on the
initial guess given for the root. However, which initial
approximation we shall take is a difficult work, especially at the
beginning. Fortunately, we can easily find the low overtone
quasinormal frequencies by means of the Hill-determinant method.

\vspace*{0.4cm}

{\bf [The Hill-determinant method]:} The condition that nontrivial
solutions of Eq. (\ref{rec}) exist is given by the requirement
that the Hill-determinant vanish \cite{Majumdar}\cite{Leaver2}.
The determinant can be easily seen in this case to be of the form
\begin{eqnarray}
 \label{max}
D= \left|
\begin{array}{llllllllll} \beta_0 & \alpha_0 & & & & & & & &\\
                      \gamma_1& \beta_1 & \alpha_1 & & & & & & &\\
                      &\gamma_2& \beta_2 & \alpha_2 & &  & & & &\\
                    &  & . & . & . &  & &  &  & \\
                        &  &  & . & . &~~  . & &  & &  \\
                            &  &  &  &. & ~~ . & . &  & & \\
                & & &   &  &\gamma_{m-1}& \beta_{m-1} & \alpha_{m-1} & &  \\
               & & &    & & &\gamma_m& \beta_m & \alpha_m & ... \\
&  &  &  &  & &  & ~.  & ~. & ~. \\
&  &  &  &  &  &  &  & ~. & ~. \\
&  &  &  &  &~~   & &  &   & ~.
\end{array} \right|
=0
 \end{eqnarray}
If $D_m$ stand for the $(m+1)\times(m+1)$ determinant of $D$, then
the $D_m$ can be concisely expressed in terms of a difference
relation \cite{Majumdar}
 \begin{eqnarray}
 D_m=\beta_m D_{m-1}-\gamma_m\alpha_{m-1}D_{m-2}.
 \end{eqnarray}
These equations are polynomials in $\omega$ and can be solved at
each order. However, the further requirement that the waves be
purely outgoing at infinity and obey the stability conditions
restrict the frequencies to complex plane with negative imaginary
parts which are just the quasinormal frequencies we look for.

The method is conceptually simpler than the continued fraction
method of Leaver. However, the method looses accuracy for the
higher overtones since one must evaluate the determinant at values
of order $N$ (for giving $N$, one require solving polynomials of
the order of $2N+2$ in $\omega$) very much higher than the
overtone number $n$ of the quasinormal modes. Also, the method
does not have any control over the stability of the root search
algorithm.

We have checked that both the methods present the same results for
$\lambda=5$ to $\lambda=10$ and $n=0$ to $n=15$.

\section{Dirac Quasinormal Modes of the Schwarzschild black hole}

In this section we represent the numerical results of the Dirac
quasinormal frequencies of the Schwarzschild black hole obtained
by using the numerical approaches just outlined in the previous
section. The results will be organized into two subsections: the
fundamental quasinormal modes and high overtones.

\subsection{Fundamental quasinormal modes}

\begin{figure}
\includegraphics[scale=1.]{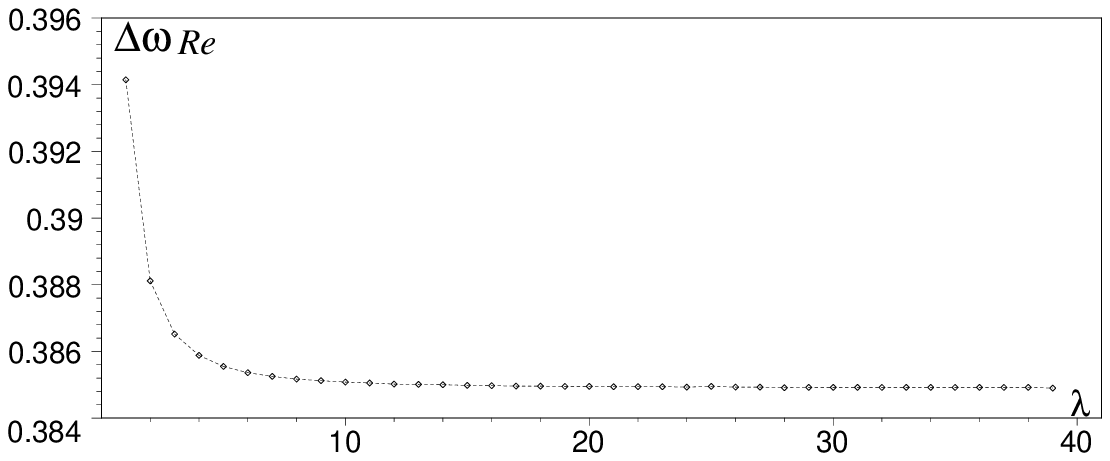}\\ \vspace*{0.3cm}
\includegraphics[scale=1.]{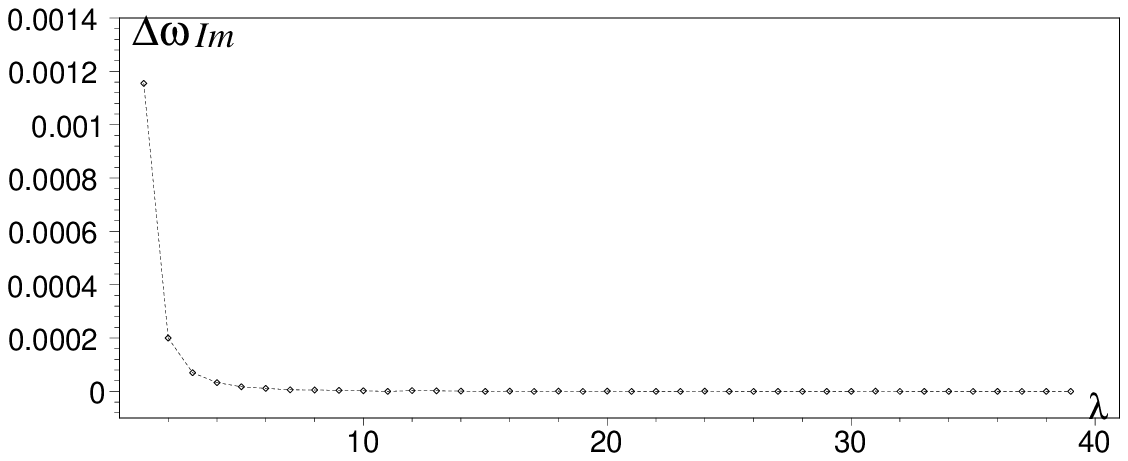}
\caption{\label{fig1} The spacing $\Delta
\omega=\omega_{\lambda+1}-\omega_{\lambda}$ as the functions of
$\lambda$ for the fundamental quasinormal frequencies. The top
figure is drawn for $\Delta \omega _{Re}
=Re(\omega_{\lambda+1})-Re(\omega_{\lambda})$ which shows that the
spacing of the real part is 0.38490 for large $\lambda$. The
bottom one is for $\Delta
\omega_{Im}=Im(\omega_{\lambda+1})-Im(\omega_{\lambda})$ which
shows that the spacing of the imaginary part becomes zero for
large $\lambda$. }
\end{figure}

The fundamental quasinormal frequencies (n=0) for $\lambda=1$ to
$\lambda=40$ are listed in the table (\ref{table1}). Fig.
(\ref{fig1}) shows that $\Delta
\omega=\omega_{\lambda+1}-\omega_{\lambda}$ as a function of
$\lambda$. From the table and figure we know that the fundamental
quasinormal frequencies become evenly spaced for large $\lambda$
and the spacing is given by
\begin{eqnarray}
\Delta \omega=0.38490-0.00000i.
\end{eqnarray}

\begin{table}
\caption{\label{table1} The fundamental Dirac quasinormal
frequencies of the Schwarzschild black hole for $\lambda=1$ to
$\lambda=40$.}
\begin{tabular}{c|c|c|c|c|c|c|c}
 \hline \hline
 ~~$\lambda$ ~~ & $\omega$  & ~~$\lambda$~~   & $\omega$ &~~$\lambda$ ~~ &
 $\omega$  & ~~$\lambda$~~   & $\omega$ \\
\hline
1 &  0.365926-0.193965i & 11& 4.23212-0.192462i& 21 & 8.08197-0.192453i & 31 & 11.93128-0.192452i\\
2 &  0.760074-0.19281i  & 12& 4.61717-0.192462i& 22 & 8.46691-0.192453i & 32 & 12.31620-0.192451i\\
3 &  1.14819-0.19261i   & 13& 5.00219-0.192459i& 23 & 8.85185-0.192453i & 33 &  12.70112-0.192451i\\
4 &  1.53471-0.19254i   & 14& 5.38720-0.192457i& 24 & 9.23679-0.192453i & 34 &  13.08604-0.192451i\\
5 &  1.92059-0.192507i & 15&  5.77220-0.192456i& 25 & 9.62172-0.192452i & 35 &  13.47096-0.192451i\\
6 &  2.30614-0.192490i & 16&  6.15718-0.192456i& 26 & 10.00667-0.192452i & 36 & 13.85588-0.192451i \\
7 &  2.69150-0.192479i & 17&  6.54215-0.192455i& 27 & 10.39160-0.192452i & 37 &  14.24080-0.192451i\\
8 &  3.07675-0.192473i & 18&  6.92711-0.192455i& 28 & 10.77653-0.192452i & 38 & 14.62572-0.192451i \\
9 &  3.46192-0.192468i & 19&  7.31207-0.192454i& 29 & 11.16144-0.192452i & 39 &  15.01064-0.192451i\\
10 & 3.84704-0.192464i & 20&  7.69702-0.192454i& 30 & 11.54636-0.192452i & 40 & 15.39554-0.192451i \\
 \hline \hline
\end{tabular}
\end{table}

\subsection{High overtones}

\begin{figure}
\includegraphics[scale=1.]{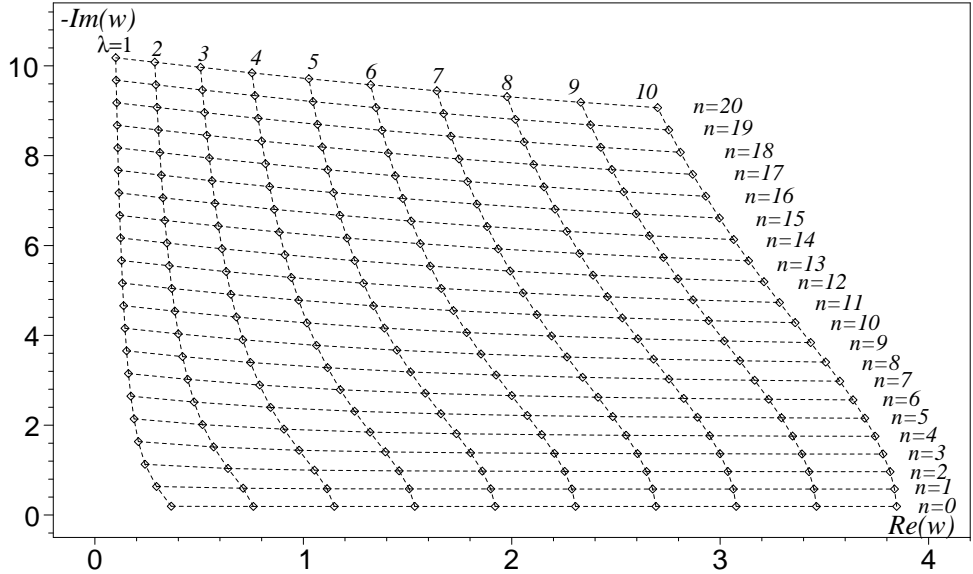}
\caption{\label{fig2} First 21 Dirac quasi-normal frequencies of
the Schwarzschild  black hole for $\lambda=1$ to $\lambda=10$. }
\end{figure}

The Dirac quasinormal frequencies of the Schwarzschild black hole
for $\lambda=1$ to $\lambda=5$ and $n=1$ to $n=20$ are given by
table (\ref{table2}) and the first 21  Dirac quasi-normal
frequencies for $\lambda=1$ to $\lambda=10$ are described by Fig.
(\ref{fig2}).

\begin{figure}
\includegraphics[scale=1.]{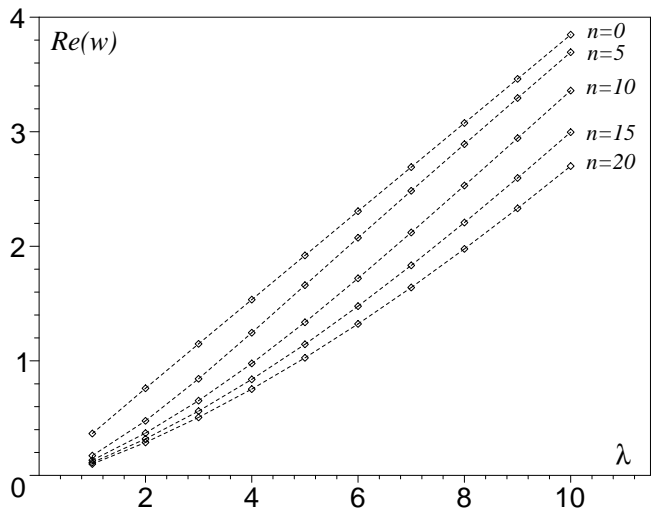}
\caption{\label{fig3} Dependence of the real parts of the
quasinormal frequencies on  $\lambda$ for the Schwarzschild black
hole, which shows that  $\lambda$ has the surprising effect of
increasing $Re(\omega)$.}
\end{figure}

\begin{figure}
\includegraphics[scale=1.]{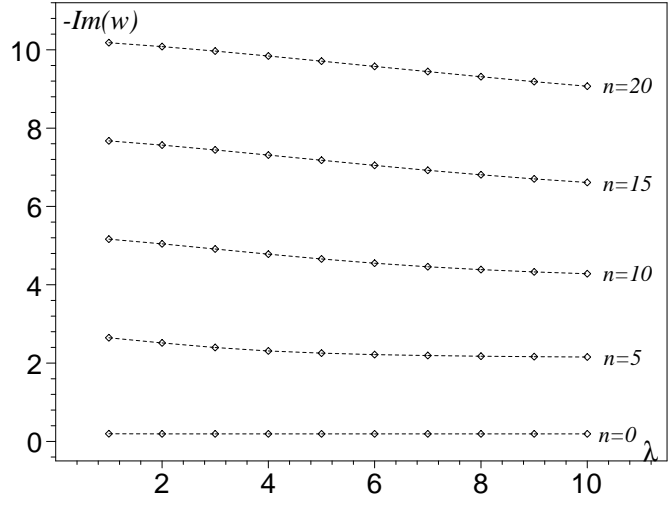}
\caption{\label{fig4} Dependence of the imaginary parts of the
quasinormal frequencies on  $\lambda$ for the Schwarzschild black
hole, which shows that  $\lambda$ almost does not affect imaginary
part, $Im(\omega)$, especially for the lower overtones.}
\end{figure}

The dependence of the quasinormal frequencies on $\lambda$ for the
Schwarzschild black hole is described by Figs. (\ref{fig3}) and
(\ref{fig4}). We learn from Figs. (\ref{fig3}) and (\ref{fig4})
that $\lambda$ has the surprising effect of increasing real part
of the quasinormal frequencies, but it almost does not affect
imaginary part, especially for the lower overtones.

\begin{figure}
\includegraphics[scale=1.]{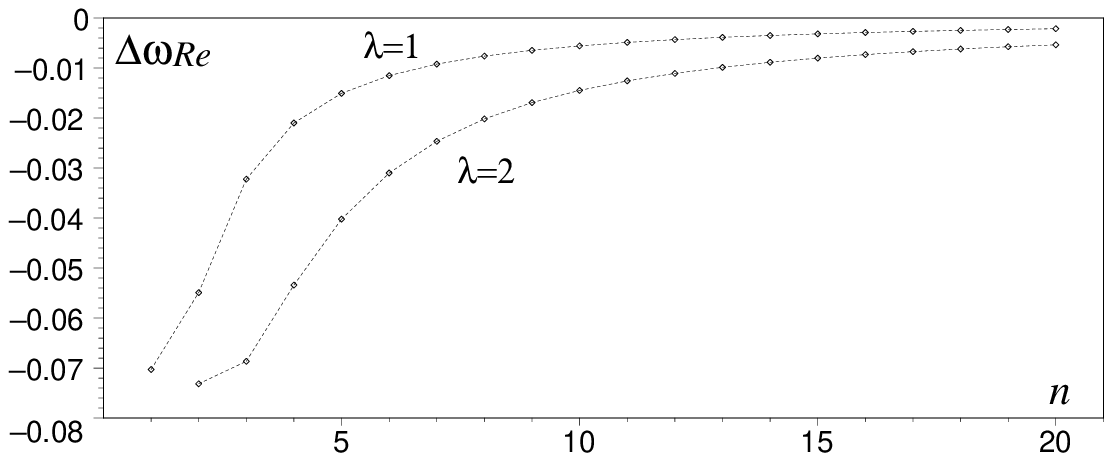}\\
\vspace*{0.5cm}
 \includegraphics[scale=1.]{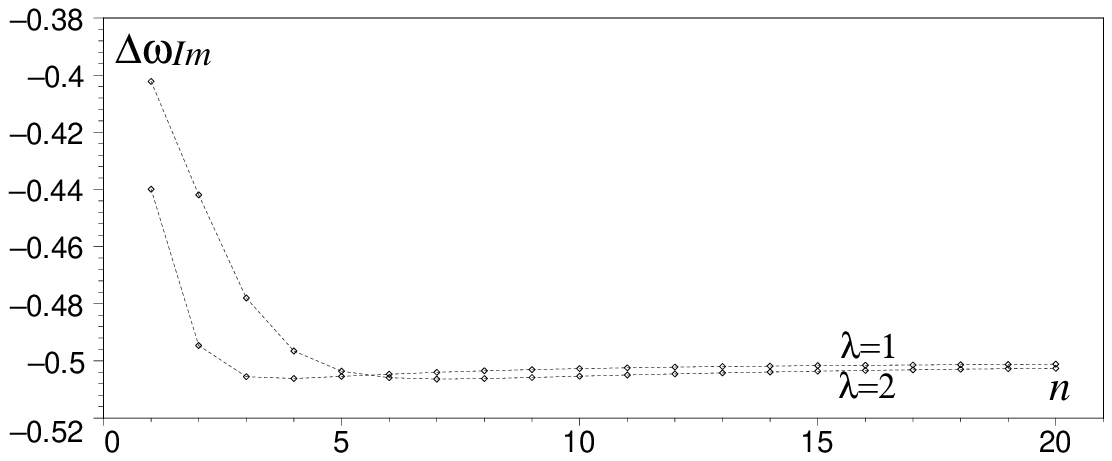}
\caption{\label{fig5} Graphs of $\Delta
\omega=\omega_{n+1}-\omega_{n}$ versus overtone number $n$ for
$\lambda=1, 2$. The top figure is drawn for $\Delta \omega _{Re}
=Re(\omega_{n+1})-Re(\omega_{n})$ which shows that $\Delta \omega
_{Re}$  tends to zero for large $n$. The bottom one is for $\Delta
\omega_{Im}=Im(\omega_{n+1})-Im(\omega_{n})$ which shows that
$\Delta \omega _{Im}\approx - \frac{i}{4M}$ for large $n$.}
\end{figure}

From table (\ref{table2}) and Fig. (\ref{fig5}) we know that the
Dirac QNMs for $\lambda=1, 2$ demonstrate the following asymptotic
behavior
 \begin{eqnarray}
&&Re(\omega_{n+1})-Re(\omega_{n})\rightarrow 0 ~~~~~~~~~
~{\text{as}}~~~n\rightarrow \infty, \\
&&Im(\omega_{n+1})-Im(\omega_{n})\approx -
\frac{i}{4M}~~~~{\text{as}}~~~n\rightarrow \infty.
 \end{eqnarray}
It is shown that the spacing for imaginary part is $-i/4M$ which
is same as that of the scalar, electromagnetic, and gravitational
perturbations.

\begin{table}
\caption{\label{table2} Dirac quasinormal frequencies of the
Schwarzschild black hole for $\lambda=1$ to $\lambda=5$ and $n=1$
to $n=20$.}
\begin{tabular}{c|c|c|c|c|c}
 \hline \hline
 n  & $\omega$~  ({$\lambda=1$)  } &  $\omega$ ~({$\lambda=2$)
  }&  $\omega$ ~ ({$\lambda=3$)  }&  $\omega$ ~ ({$\lambda=4$)  }
   &  $\omega$ ~ ({$\lambda=5$)  }\\
\hline 1 & 0.295644-0.633857i &  0.711666-0.594995i  &
1.11403-0.58543i
 &  1.50860-0.581936i  &  1.89952-0.580296i
          \\
2 & 0.240740-1.128450i &   0.638523-1.03691i  &  1.05321-0.99939i
 &  1.45954-0.98382i  &  1.85899-0.976231i
           \\
3 & 0.208512-1.63397i  &  0.569867-1.5149i    & 0.97949-1.44176i
 & 1.39383-1.40459i   &  1.80226-1.38504i
          \\
4 & 0.187528-2.14013i  &  0.516459-2.01144i   & 0.906879-1.91029i
 &  1.3199-1.84711i   &  1.73409-1.81009i
             \\
5 & 0.17248-2.64556i   & 0.476221-2.51504i    & 0.843213-2.3969i
 & 1.24587-2.3100i    & 1.66006-2.25269i
             \\
6 & 0.160962-3.15021i  &  0.445224-3.0209i    & 0.790087-2.89386i
 & 1.17721-2.78892i   &  1.58544-2.71190i
             \\
7 & 0.151745-3.65419i  &  0.420575-3.52725i   & 0.746214-3.39616i
 & 1.11627-3.27894i   &  1.51418-3.18513i
             \\
8 & 0.144128-4.15764i  &  0.400401-4.03343i   & 0.70972-3.90102i
 & 1.06324-3.77605i   &  1.44849-3.66916i
              \\
9 & 0.13768-4.66067i   & 0.383487-4.53924i   & 0.678957-4.40701i
 &  1.01733-4.27745i  & 1.38920-4.16097i
              \\
10 & 0.132118-5.16336i &  0.369026-5.0446i   &  0.65265-4.91341i
 &  0.97747-4.78135i  &  1.33621-4.65815i
             \\
11 & 0.12725-5.66576i  &  0.356461-5.54954i  & 0.629847-5.41982i
 &  0.942637-5.28666i &  1.28901-5.15892i
             \\
12 & 0.122937-6.16792i &  0.345399-6.05409i  & 0.60984-5.92607i
 &  0.911955-5.79270i &  1.24690-5.66202i
             \\
13 & 0.119077-6.66989i &  0.335551-6.55828i  & 0.59210-6.43207i
 &  0.884707-6.29908i &  1.20921-6.16662i
              \\
14 & 0.115592-7.17170i  &  0.326701-7.06215i  & 0.57622-6.93778i
 &  0.860321-6.80556i &  1.17532-6.67215i
              \\
15 & 0.112424-7.67335i &  0.318685-7.56574i  & 0.56189-7.44319i
 & 0.838336-7.31197i  &  1.14467-7.17822i
              \\
16 & 0.109525-8.17489i &  0.311372-8.06907i & 0.548867-7.94831i &
0.818386-7.81825i & 1.11682-7.68457i
\\
17 & 0.106858-8.67631i &   0.304662-8.57218i & 0.536959-8.45316i &
0.800173-8.32435i & 1.09138-8.19105i
\\
18 & 0.104393-9.17764i &  0.298471-9.07509i &  0.526008-8.95774i &
0.783457-8.83023i & 1.06803-8.69753i
\\
19 & 0.102104-9.67888i &   0.292733-9.57782i &  0.515889-9.46207i
& 0.768038-9.33589i  & 1.04651-9.20394i
\\
20 & 0.099979-10.1800i  &  0.287392-10.0804i &  0.506497-9.96618i
& 0.753754-9.84132i & 1.02661-9.71027i
\\
 \hline \hline
\end{tabular}
\end{table}

\section{summary}

The wave equations for the Dirac fields in the Schwarzschild black
hole spacetime are obtained by means of the Newman-Penrose
formulism. Then, the quasinormal frequencies corresponding to the
Dirac field perturbation in the black hole spacetime are evaluated
by using continued fraction and Hill-determinant approaches. Three
interesting results obtained in this paper are listed in the
following: (i) The fundamental quasinormal frequencies become
evenly spaced for large $\lambda$ and the spacing is given by
$\omega_{\lambda+1}-\omega_{\lambda}=0.38490-0.00000i$. (ii)
$\lambda$ has the surprising effect of increasing real part of the
quasinormal frequencies, but it almost does not affect imaginary
part, especially for the low overtones. (iii) The quasinormal
frequencies also become evenly spaced for large $n$ and the
spacing for imaginary part is $-i/4M$, which is same as that of
the scalar, electromagnetic, and gravitational perturbations.

In future papers in this series the new wave functions and
potentials will be applied to study of Dirac QNMs in other black
hole spacetimes, such as the charged Reissner-Norstr\"om black
hole and the rotating Kerr black hole.

\begin{acknowledgments}This work was supported by the
National Natural Science Foundation of China under Grant No.
10275024 and under Grant No. 10473004; the FANEDD under Grant No.
200317; and the SRFDP under Grant No. 20040542003.
\end{acknowledgments}

\end{document}